\documentclass{JHEP3}

\usepackage{epsfig}
\usepackage{amsmath}
\usepackage{amssymb,amsfonts}
\usepackage{graphicx}
\usepackage{graphics}
\usepackage{amsthm}
\usepackage{graphicx}
\usepackage{multirow}
\usepackage{subfigure}
\newcommand{\cA}{{\cal A}}
\newcommand{\cAb}{{\overline{\cal A}}}
\newcommand{\cF}{{\cal F}}
\newcommand{\cFb}{{\overline{\cal F}}}
\newcommand{\cD}{{\cal D}}
\newcommand{\cDb}{{\overline{\cal D}}}
\newcommand{\cQ}{{\cal Q}}
\newcommand{\cU}{{\cal U}}
\newcommand{\cN}{{\cal N}}
\newcommand{\cUb}{{\overline{\cal U}}}
\newcommand{\Tr}{{\rm Tr\;}}

\newcommand{\vn}{ {\bf n} }
\newcommand{\KD}{{K\"{a}hler--Dirac }}
\newcommand{\hatbmu}{\widehat{\boldsymbol {\mu}}}
\newcommand{\hatbe}{\widehat{\boldsymbol {e}}}

\def\nn{\nonumber}
\def\bec{\begin{center}}
\def\eec{\end{center}}
\def\beq{\begin{equation}}
\def\eeq{\end{equation}}
\def\bea{\begin{eqnarray}}
\def\eea{\end{eqnarray}}

\title{On the sign problem in 2D lattice super Yang--Mills}

\author{Simon Catterall \\
Department of Physics, Syracuse University, Syracuse, NY13244, USA
}

\author{Richard Galvez \\
Department of Physics, Syracuse University, Syracuse, NY13244, USA
}

\author{Anosh Joseph \\
Theoretical Division, Los Alamos National Laboratory, Los Alamos, NM 87545, USA
}

\author{Dhagash Mehta \\
Department of Physics, Syracuse University, Syracuse, NY13244, USA
}

\preprint{LA-UR-11-11773}

\abstract{In recent years a new class of supersymmetric lattice theories have been proposed which retain one or more exact supersymmetries for non-zero lattice spacing. Recently there has been some controversy in the literature concerning whether these theories suffer from a sign problem. In this paper we address this issue by conducting simulations of the $\cN=(2, 2)$ and $\cN=(8, 8)$ supersymmetric Yang--Mills theories in two dimensions for the $U(N)$ theories with $N=2,3,4$, using the new twisted lattice formulations. Our results provide evidence that these theories do {\it not} suffer from a sign problem in the continuum limit. These results thus boost confidence that the new lattice formulations can be used successfully to explore non-perturbative aspects of four-dimensional $\cN=4$ supersymmetric Yang--Mills theory.}


\begin{document}

\section{Introduction}

Supersymmetric Yang-Mills (SYM) theories are interesting from a variety of perspectives; as toy models for understanding theories such as QCD, as potential theories of BSM physics and via the AdS/CFT correspondence because of a possible connection to quantum gravity. Many features of these theories, for example, dynamical supersymmetry breaking, are inherently non-perturbative in nature and this serves as motivation to study such theories on the lattice.

Unfortunately, historically it has proven difficult to discretize supersymmetric theories using traditional methods. This stems from the fact that the supersymmetry algebra is an extension of the usual Poincar\'e algebra and hence is broken completely by na\"ive discretization on a space-time lattice. However, recently the development of a series of new theoretical tools have enabled us to construct certain supersymmetric theories on the lattice while preserving a subset of the continuum supersymmetries - see the reviews \cite{Kaplan:2003uh, Giedt:2006pd, Catterall:2009it, arXiv:1110.5983} and references therein. Other recent complementary approaches to the problem of exact lattice supersymmetry can be found in \cite{Sugino:2003yb, Sugino:2004qd, hep-lat/0507029, arXiv:0707.3533, Kanamori:2008bk, Hanada:2009hq, Hanada:2010kt, Hanada:2010gs, Hanada:2011qx}.

One way to understand the new constructions is to realize that they correspond to discretizations of topologically twisted forms of the target continuum theories. Currently, lattice constructions exist for a set of SYM theories, including the four-dimensional $\cN=4$ SYM theory.

Lattice theories constructed this way are free of doublers, respect gauge-invariance, preserve a subset of the original supersymmetries and target the usual continuum theories in the na\"ive continuum limit. These constructions are possible only if the continuum SYM theories possess sufficient extended supersymmetry; the precise requirement is that the number of supercharges must be an integer multiple of $2^D$ where $D$ is the space-time dimension. This includes the $\cN=(2, 2)$ SYM theory in two dimensions and $\cN = 4$ SYM in four dimensions. In this paper we study both theories in two dimensions -the $\cN=4$ model
 yielding the $\cN=(8,8)$ theory after dimensional reduction from four to two dimensions.

However, even when a supersymmetric lattice construction exists, it is still possible to encounter an additional difficulty that renders the use of numerical simulation problematic -- the fermionic sign problem. To understand the nature of this problem consider a generic lattice theory with a set of bosonic $\phi$ and fermionic $\psi$ degrees of freedom. The partition function of the theory is
\bea
Z &=& \int [d\phi][d\psi]\, \exp\Big(-S_B[\phi] - \psi^T M[\phi] \psi\Big)~,\nn \\
&=& \int [d\phi]\, {\rm Pf}(M)~\exp \Big(-S_B[\phi]\Big)~,
\eea
where $M$ is antisymmetric fermion matrix and ${\rm Pf}(M)$ the corresponding Pfaffian. For a $2n \times 2n$ matrix $M$, the Pfaffian is explicitly given as ${\rm Pf}(M)^{2} = \mbox{Det}\, M$. In the supersymmetric lattice constructions we will consider in this paper, $M$ at non zero lattice spacing is a complex operator and one might worry that the resulting Pfaffian could exhibit a fluctuating phase depending on the background boson fields $\phi$. Since Monte Carlo simulations must be performed with a positive definite measure, the only way to incorporate this phase is through a reweighting procedure, which folds this phase in with the observables of the theory. Expectation values of observables derived from such simulations can then suffer drastic statistical errors which overwhelm the signal -- the famous fermionic {\it sign problem}. Thus, if such a complex phase is present, the Monte Carlo technique is rendered effectively useless. Lattice theories such as QCD with finite chemical potential are known to suffer from a severe sign problem, which makes it very difficult to extract physical observables from simulations using conventional methods. The lattice sign problem exists not only in relativistic field theories but also in a variety of condensed matter systems \cite{Hirsch:1983rq}.

In the construction of supersymmetric lattice gauge theories, there has been an ongoing debate on the existence of a sign problem in the two-dimensional $\cN=(2,2)$ supercharge lattice theory \cite{Giedt:2003ve, Catterall:2008dv, Hanada:2010qg}. The resolution of this sign problem is crucial as the extraction of continuum physics from the lattice model depends very much on whether the results from phase quenched simulations can be trusted. Moreover, if a sign problem were to be found in this model it makes it more likely that the four-dimensional $\cN = 4$ theory also suffers from a sign problem which would render practical simulation of this theory impossible. In \cite{Giedt:2003ve}, it was shown that there is a potential sign problem in the two-dimensional $\cN = (2, 2)$ SYM lattice theory. Furthermore, in \cite{Catterall:2008dv} numerical evidence was presented of a sign problem in a phase quenched dynamical simulation of the theory at non-zero lattice spacing. More recently Hanada et al. \cite{Hanada:2010qg} have argued that there is no sign problem for this theory in the continuum limit. However, the models studied by these various groups differed in detail; Catterall et al. studied an $SU(2)$ model obtained by truncating the supersymmetric $U(2)$ theory and utilized bosonic link fields valued in the group $SL(2,C)$, while Hanada et al. used a $U(2)$ model where the complexified bosonic variables take their values in the algebra of $U(2)$ together with the inclusion of supplementary mass terms to control scalar field fluctuations.

In this paper, we present results from simulations of the two dimensional $\cN=(2, 2)$ $U(N)$ SYM theory (which we will refer to from now on as the $\cQ = 4$ theory, with $\cQ$ the number of supercharges) and the maximally supersymmetric $\cN = (8, 8)$ $U(N)$ SYM theory (we refer to this theory as the $\cQ = 16$ theory). Our results provide strong
evidence that there is no sign problem in the supersymmetric continuum limit for these theories.
In the next four sections we summarize the details of the lattice constructions of both theories including a discussion of the possible parameterizations of the bosonic link fields. We then present our numerical results for $\cQ = 4$ and $\cQ = 16$ lattice SYM theories in two dimensions.

\section{Supersymmetric Yang--Mills theories on the lattice}

As discussed in the introduction it is possible to discretize a class of continuum SYM theories using ideas based on topological twisting\footnote{Note that the lattice actions constructed using orbifold and twisted methods are equivalent \cite{Unsal:2006qp, Catterall:2007kn, Damgaard:2007xi}.}. Though the basic idea of twisting goes back to Witten in his seminal paper on topological field theory \cite{Witten:1988ze}, it actually had been anticipated in earlier work on staggered fermions \cite{Elitzur:1982vh}. In our context, the idea of twisting is to decompose the fields of the Euclidean SYM theory in $D$ space-time dimensions in representations not in terms of the original (Euclidean) rotational symmetry $SO_{\rm rot}(D)$, but a twisted rotational symmetry, which is the diagonal subgroup of this symmetry and an $SO_{\rm R}(D)$ subgroup of the R-symmetry of the theory, that is,
\beq
SO(D)^\prime={\rm diag}(SO_{\rm Lorentz}(D)\times SO_{\rm R}(D))~.
\eeq
As an example, let us consider the case where the total number of supersymmetries is $Q=2^D$. In this case we can treat the supercharges of the twisted theory as a $2^{D/2}\times 2^{D/2}$ matrix $q$. This matrix can be expanded on the Dirac--K\"ahler basis as
\beq
q = \cQ I + \cQ_a \gamma_a + \cQ_{ab}\gamma_a\gamma_b + \ldots
\eeq
The $2^D$ antisymmetric tensor components that arise in this basis are the twisted supercharges that satisfy the corresponding supersymmetry algebra inherited from the original algebra
\bea
\cQ^2 &=& 0\\
\{\cQ,\cQ_a\} &=& p_a\\
&\vdots&
\eea
The presence of the nilpotent scalar supercharge $\cQ$ is most important; it is the algebra of this charge that is compatible with discretization. The second piece of the algebra expresses the fact that the momentum is the $\cQ$-variation of something which makes the statement plausible that the energy-momentum tensor and hence the entire action can be written in a $\cQ$-exact form\footnote{In the case of four-dimensional $\cN = 4$ SYM there is an additional $\cQ$-closed term in the action.}. Notice that an action written in such a $\cQ$-exact form is trivially invariant under the scalar supersymmetry $\cQ$ provided the latter remains nilpotent under discretization.

The recasting of the supercharges in terms of twisted variables can be repeated for the fermions of the theory and yields a set of antisymmetric tensors $(\eta, \psi_a, \chi_{ab}, \ldots)$, which for the case of $Q=2^D$ matches the number of components of a real \KD field. This repackaging of the fermions of the theory into a \KD field is at the heart of how the discrete theory avoids fermion doubling as was shown by Becher, Joos and Rabin in the early days of lattice gauge theory \cite{Rabin:1981qj, Becher:1982ud}. It is important to recognize that the transformation to twisted variables corresponds to a simple change of variables in flat space -- one more suitable for discretization.

\subsection{Two-dimensional $\cQ=4$ SYM on the lattice}
\label{sec:2d-formulation}

The two-dimensional $\cQ = 4$ SYM theory is the simplest example of a gauge theory that permits topological twisting and thus satisfies our requirements for supersymmetric lattice constructions. Its R-symmetry possesses an $SO(2)$ subgroup corresponding to rotations of the its two degenerate Majorana fermions into each other. After twisting the fields and supersymmetries of the target theory, the action takes the following form in the continuum
\beq
S = \frac{1}{g^2} \cQ \int \Tr \left(\chi_{ab}\cF_{ab} + \eta [ \cDb_a,\cD_b ] - \frac{1}{2}\eta d\right)~,
\label{2daction-twisted}
\eeq
where $g$ is the coupling parameter. We use an anti-hermitian basis for the generators of the gauge group with ${\rm Tr}(T^a T^b)=-\delta^{ab}$.

The degrees of freedom appearing in the above action are just the twisted fermions $(\eta, \psi_a, \chi_{ab})$ and a complexified gauge field $\cA_a$. The latter is built from the usual gauge field $A_a$ and the two scalars $B_a$ present in the untwisted theory: $\cA_a = A_a + iB_a$. The twisted theory is naturally written in terms of the complexified covariant derivatives
\beq
\cD_a = \partial_a + \cA_a,~~~\cDb_a = \partial_a + \cAb_a~,
\eeq
and complexified field strengths
\beq
\cF_{ab} = [\cD_a, \cD_b],~~~\cFb_{ab} = [\cDb_a, \cDb_b]~.
\eeq
Notice that the original scalar fields transform as vectors under the original R-symmetry and hence become vectors under the twisted rotation group while the gauge fields are singlets under the R-symmetry and so remain vectors under twisted rotations. This structure makes the appearance of a complex gauge field in the twisted theory possible. This action is invariant under the original $U(N)$ gauge symmetry from the untwisted theory.

The nilpotent transformations associated with the scalar supersymmetry $\cQ$ are given explicitly by
\bea
\cQ\; \cA_a &=& \psi_a \nn \\
\cQ\; \psi_a &=& 0 \nn \\
\cQ\; \cAb_a &=& 0 \nn \\
\cQ\; \chi_{ab} &=& -\cFb_{ab} \nn \\
\cQ\; \eta &=& d \nn \\
\cQ\; d &=& 0
\eea

Performing the $\cQ$-variation on the action and integrating out the auxiliary field $d$ yields
\beq
S = \frac{1}{g^2} \int \Tr \left(-\cFb_{ab}\cF_{ab} + \frac{1}{2}[ \cDb_a, \cD_a]^2 - \chi_{ab}\cD_{\left[a\right.}\psi_{\left.b\right]}-\eta \cDb_a\psi_a\right)~.
\label{2d-twist_action}
\eeq

\begin{figure}
\begin{center}\includegraphics[width=0.5\textwidth]{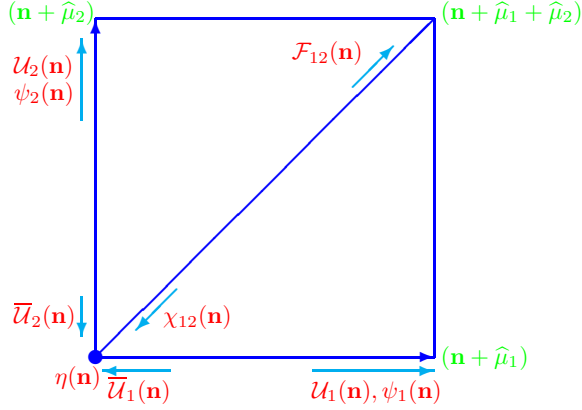}\end{center}
\caption{\label{fig:2dlattice}The 2d lattice for the four supercharge theory with field orientation assignments.}
\end{figure}

The prescription for discretization is somewhat natural. The complexified gauge fields are represented as complexified Wilson gauge fields
\beq
\cA_a(x) \rightarrow \cU_a(\vn)~,
\eeq
living on links of a lattice, which for the moment can be thought of as hypercubic, with integer-valued basis vectors
\beq
\hatbmu_1 = (1, 0),~~~\hatbmu_2 = (0, 1)~.
\eeq
They transform in the usual way under $U(N)$ lattice gauge transformations
\beq
\cU_a(\vn)\to G(\vn)\cU_a(\vn)G^\dagger(\vn+\hatbmu_a)~.
\eeq
Supersymmetric invariance then implies that $\psi_a(\vn)$ live on the same links and transform identically. The scalar fermion $\eta(\vn)$ is clearly most naturally associated with a site and transforms accordingly
\beq
\eta(\vn)\to G(\vn)\eta(\vn)G^\dagger(\vn)~.
\eeq
The field $\chi_{ab}(\vn)$ is slightly more difficult. Naturally as a 2-form it should be associated with a plaquette. In practice we introduce diagonal links running through the center of the plaquette and choose $\chi_{ab}(\vn)$ to lie {\it with opposite orientation} along those diagonal links. This choice of orientation will be necessary to ensure gauge invariance. Figure \ref{fig:2dlattice} shows the resultant lattice theory.

To complete the discretization we need to describe how continuum derivatives are to be replaced by difference operators. A natural technology for accomplishing this in the case of adjoint fields was developed many years ago and yields expressions for the derivative operator applied to arbitrary lattice p-forms \cite{Aratyn:1984bd}. In the case discussed here we need just two derivatives given by the expressions
\bea
\cD^{(+)}_a f_b(\vn) &=& \cU_a(\vn)f_b(\vn + \hatbmu_a) - f_b(\vn)\cU_a(\vn+ \hatbmu_b)~,\\
\cDb^{(-)}_a f_a(\vn) &=& f_a(\vn)\cUb_a(\vn)-\cUb_a(\vn - \hatbmu_a)f_a(\vn - \hatbmu_a)~.
\eea
The lattice field strength is then given by the gauged forward difference acting on the link field: $\cF_{ab}(\vn) = \cD^{(+)}_a \cU_b(\vn)$, and is automatically antisymmetric in its indices. Furthermore, it transforms like a lattice 2-form and yields a gauge invariant loop on the lattice when contracted with $\chi_{ab}(\vn)$. Similarly the covariant backward difference appearing in $\cDb^{(-)}_a \cU_a(\vn)$ transforms as a 0-form or site field and hence can be contracted with the site field $\eta(\vn)$ to yield a gauge invariant expression.

This use of forward and backward difference operators guarantees that the solutions of the lattice theory map one-to-one with the solutions of the continuum theory and hence fermion doubling problems are evaded \cite{Rabin:1981qj}. Indeed, by introducing a lattice with half the lattice spacing one can map this \KD fermion action into the action for staggered fermions \cite{Banks:1982iq}. Notice that, unlike the case of QCD, there is no rooting problem in this supersymmetric construction since the additional fermion degeneracy is already required in the continuum theory.

As for the continuum theory the lattice action is again $\cQ$-exact:
\beq
S = \sum_{\vn} \Tr \cQ \Big(\chi_{ab}(\vn)\cD_a^{(+)}\cU_b(\vn) + \eta(\vn) \cDb_a^{(-)}\cU_a(\vn) - \frac{1}{2}\eta(\vn) d(\vn) \Big)~.
\eeq
Acting with the $\cQ$ transformation on the lattice fields and integrating out the auxiliary field $d$, we obtain the gauge and $\cQ$-invariant lattice action:
\beq
\label{eq:2d-latt-action}
S = \sum_{\vn} \Tr \Big(\cF_{ab}^{\dagger}(\vn) \cF_{ab}(\vn) + \frac{1}{2}\Big(\cDb_a^{(-)}\cU_a(\vn)\Big)^2 - \chi_{ab}(\vn) \cD^{(+)}_{[a}\psi_{b]}(\vn) - \eta(\vn) \cDb^{(-)}_a\psi_a(\vn) \Big)~.
\eeq

\subsection{Four-dimensional $\cQ = 16$ SYM on the lattice}
\label{sec:4d-lattice-theory}

In four dimensions the constraint that the target theory possess sixteen supercharges singles out a unique theory for which this construction can be undertaken -- the $\cN = 4$ SYM theory.

The continuum twist of $\cN = 4$ that is the starting point of the twisted lattice construction was first written down by Marcus in 1995 \cite{Marcus:1995mq} although it now plays an important role in the Geometric-Langlands program and is hence sometimes called the GL-twist \cite{Kapustin:2006pk}. This four-dimensional twisted theory is most compactly expressed as the dimensional reduction of a five-dimensional theory in which the ten (one gauge field and six scalars) bosonic fields are realized as the components of a complexified five-dimensional gauge field while the 16 twisted fermions naturally span one of the two \KD fields needed in five dimensions. Remarkably, the action of this theory contains a $\cQ$-exact term of precisely the same form as the two-dimensional theory given in Eq. (\ref{2daction-twisted}) provided one extends the indices labeling the fields to run now from one to five. In addition, the Marcus twist
of $\cN=4$ YM requires a new $\cQ$-closed term which was not possible in the two-dimensional theory
\beq
S_{\rm closed} = -\frac{1}{8} \int \Tr \epsilon_{mnpqr} \chi_{qr} \cDb_p \chi_{mn}~.
\label{closed}
\eeq
The supersymmetric invariance of this term then relies on the Bianchi identity
\beq
\epsilon_{mnpqr}\cDb_p\cFb_{qr} = 0~.
\eeq

The four-dimensional lattice that emerges from examining the moduli space of the resulting discrete theory is called the $A_4^*$-lattice and is constructed from the set of five basis vectors $\hatbe_a$ pointing out from the center of a four-dimensional equilateral simplex out to its vertices together with their inverses $ -\hatbe_a$. It is the four-dimensional analog of the two-dimensional triangular lattice. Complexified Wilson gauge link variables $\cU_a$ are placed on these links together with their $\cQ$-superpartners $\psi_a$. Another 10 fermions are associated with the diagonal links $\hatbe_a + \hatbe_b$ with $a>b$. Finally, the exact scalar supersymmetry implies the existence of a single fermion for every lattice site. The lattice action corresponds to a discretization of the Marcus twist on this $A_4^*$-lattice and can be represented as a set of traced closed bosonic and fermionic loops. It is invariant under the exact scalar supersymmetry $\cQ$, lattice gauge transformations and a global permutation symmetry $S^5$ and can be proven free of fermion doubling problems as discussed above. The $\cQ$-exact part of the lattice action is again given by Eq. (\ref{eq:2d-latt-action}) where the indices $a, b$ now correspond to the indices labeling the five basis vectors of $A_4^*$.

While the supersymmetric invariance of this $\cQ$-exact term is manifest in the lattice theory, it is not clear how to discretize the continuum $\cQ$ closed term. Remarkably, it is possible to discretize Eq. (\ref{closed}) in such a way that it is indeed exactly invariant under the twisted supersymmetry
\beq
S_{\rm closed} = -\frac{1}{8}\sum_{\vn} \Tr \epsilon_{mnpqr} \chi_{qr}(\vn + \hatbmu_m + \hatbmu_n + \hatbmu_p)
\cDb^{(-)}_p\chi_{mn}(\vn + \hatbmu_p)
\eeq
and can be seen to be supersymmetric since the lattice field strength satisfies an exact Bianchi identity \cite{Aratyn:1984bd}.
\beq
\epsilon_{mnpqr}\cDb^{(+)}_p\cFb_{qr} = 0~.
\eeq

The renormalization of this theory has been recently studied in perturbation theory with some remarkable conclusions \cite{Catterall:2011pd}; namely that the classical moduli space is not lifted to all orders in the coupling, that the one loop lattice
beta function vanishes and that no fine tuning of the bare lattice parameters with cut-off is required at one-loop for the theory to recover full supersymmetry as the lattice spacing is sent to zero.

\section{Towards the continuum limit}
\subsection{Parametrizations of the gauge links}
\label{sec:gauge-link-params}

There exist two distinct parameterizations of the gauge fields on the lattice that
have been proposed for these theories. The first one follows the standard Wilson prescription where the complexified gauge fields in the continuum are mapped to link fields $\cU_a(\vn)$ living on the link between $\vn$ and $\vn + \hatbmu_a$ through the mapping
\beq
\cU_a(\vn) = e^{\cA_a(\vn)}~,
\eeq
where $\cA_a (\vn) = \sum_{i=1}^{N_G} \cA_a^i T^i$ and $T^i=1, \ldots, N_G$ are the anti-hermitian generators of $U(N)$.
The resultant gauge links belong to $GL(N,C)$.
We call this realization of the bosonic links the {\it exponential or group based parametrization}\footnote{Notice that our lattice gauge fields are dimensionless and hence contain an implicit factor of the lattice spacing $a$.}.

The other parametrization of the bosonic link fields that has been used, particularly in the orbifold literature, simply takes the complex gauge links as taking values in the algebra of the $U(N)$ group
\beq
\cU_a(\vn)=\cA_a(\vn)~.
\eeq
In this case to obtain the correct continuum limit one must subsequently expand the fields around a particular point in the moduli space of the theory corresponding to giving an expectation value to a component of the link field proportional to the unit matrix.  This
field can be identified as the trace mode of the scalar field in the untwisted
theory.
\beq
\cU_a(\vn) = {\mathbf I}_N + \cA_a(\vn)~.
\label{alg}
\eeq
Usually the use of
such an algebra based or {\it non compact} parametrization would signal a breaking of lattice gauge invariance. It is only possible here because the bosonic fields take values in a complexified $U(N)$ theory -- so that the unit matrix appearing in Eq. (\ref{alg}) can be interpreted as the expectation value of a {\it dynamical field} - the trace mode of the scalars. We will refer to this parametrization as the {\it linear or algebra based parametrization}\footnote{{In fact, a non-compact parametrization of the gauge-fields has also been recently used to restore  BRST symmetry on the lattice in Ref. \cite{arXiv:0710.2410}, i.e., to evade the so-called Neuberger $0/0$ problem \cite{Print-86-0394} (see also Refs.~\cite{arXiv:0710.2410} and \cite{arXiv:0912.0450} for the recent progress, and \cite{Mehta:latt11} for the relation between the Neuberger $0/0$ problem and sign problem for the lattice SYM theories.).}}.

Both parameterizations of the gauge links are equivalent at leading order in the lattice spacing, yield the same lattice action and can be considered as providing equally valid representations of the lattice theory at the classical level. The exponential parametrization was used in studies of both $\cQ = 4$ and $\cQ = 16$ theories in \cite{Catterall:2008dv} while in \cite{Hanada:2010qg} the linear parametrization was employed to perform simulations of the $\cQ = 4$ theory. In this work we have concentrated on the linear parametrization principally because it is naturally associated with a manifestly
supersymmetric measure in the path integral - the flat measure. Explicit comparison with results from the exponential parametrization
can be found in \cite{Mehta:latt11}.

\subsection{Potential terms}
\label{sec:pot-terms}

As we have described in the previous section, the linear parameterization only yields the correct na\"ive continuum limit if the trace mode of the scalars develops a vacuum expectation value so that appropriate kinetic terms are generated in the tree level action. In addition, we require that the fluctuations of all dimensionless lattice fields vanish as the lattice spacing is sent to zero; a non-trivial issue in theories possessing flat directions associated with extended supersymmetry. Since no classical scalar potential is present in the lattice theory\footnote{Lattice theories based on supersymmetric mass deformations have also been proposed in two dimensions \cite{Hanada:2010qg, Hanada:2011qx}} it is crucial to add {\it by hand} a suitable gauge invariant potential to ensure these features\footnote{It was precisely this requirement that led to a truncation of the $U(N)$ symmetry to $SU(N)$ in the original simulations of these theories. One can think of this truncation as  corresponding to the use of a delta function potential for the $U(1)$ part of the field \cite{Catterall:2008dv}.}. 
Specifically we add a potential term of the following form \cite{Hanada:2010qg}
\beq
S_M = \mu^2 \sum_\vn \left(\frac{1}{N}{\rm Tr}(\cU_a^\dagger(\vn) \cU_a(\vn))-1\right)^2~,
\eeq
to  the lattice action. Here $\mu$ is a tunable mass parameter, which can be used to control the expectation values and fluctuations of the lattice fields. Notice that such a potential obviously breaks supersymmetry -- however because of the exact supersymmetry at $\mu = 0$ all supersymmetry breaking counterterms induced via quantum effects will possess couplings that vanish as $\mu \to 0$ and so can be removed by sending $\mu\to 0$ at the end of the calculation.

To understand the effect of this term let us consider the full set of vacuum equations for the lattice theory. These are given by setting the bosonic action to zero
\bea
\cF_{ab}(\vn) &=&0~, \\
\cDb_a^{(-)}\cU_a(\vn) &=&0~, \\
\frac{1}{N}{\rm Tr}\Big(\cU^\dagger_a(\vn)\cU_a(\vn)\Big) - 1 &=& 0~.
\eea
The first two equations imply that the moduli space consists of constant complex matrices taking values in the $N$-dimensional Cartan subalgebra of $U(N)$. 

Assuming that the matrix valued complexified link fields $\cU_a(\vn)$ are nonsingular\footnote{Having zero eigenvalues for the matrices $\cU_a(\vn)$ would not cause a  problem for us, as we are interested in expanding these fields around the point ${\mathbf I}_N$ instead of the origin of the moduli space.}, we can decompose them in the following way
\beq
\cU_a(\vn) = P_a(\vn)U_a(\vn)~,
\eeq 
where $P_a(\vn)$ is a positive semidefinite hermitian matrix and $U_a(\vn)$ a unitary matrix. The form of the mass term clearly does not depend on the unitary piece and clearly is minimized by setting $P_a(\vn) = {\mathbf I}_N$. Expanding about this configuration gives the following expression for the complex link matrices
\beq
\cU_a(\vn) = P_a(\vn)U_a(\vn) = \Big({\mathbf I}_N + p_a(\vn)\Big)U_a(\vn)~,
\eeq
where $p_a(\vn)$ is a hermitian matrix.
Minimizing the mass term leads to
\bea
0 &=& \frac{1}{N}{\rm Tr}\Big(\cU^{\dagger}_a(\vn)\cU_a(\vn)\Big) - 1~,\nn \\
&=& \frac{1}{N}{\rm Tr}\Big[U^\dagger_a(\vn)\Big({\mathbf I}_N + p_a(\vn)\Big)\Big]\Big[\Big({\mathbf I}_N + p_a(\vn)\Big)U_a(\vn)\Big] - 1~,\nn \\
&=& \frac{1}{N}{\rm Tr}\Big[{\mathbf I}_N + 2p_a(\vn)+ p^2_a(\vn)\Big] - 1~,\nn \\
&=& \frac{1}{N}\Big[\frac{2}{\sqrt{N}} p^0_a(\vn)+ \sum_{A=1}^N (p^A_a(\vn))^2\Big]~.
\eea
where we have adopted a basis in which $T^0$ is proportional to the unit matrix and all
other (Cartan) generators are traceless.
Analyzing the gauge transformation properties of the complexified link fields,
\beq
\cU_a(\vn) \rightarrow G(\vn)\cU_a(\vn)G^\dagger(\vn + \hatbmu_a)~,
\eeq
we see that the unitary piece $U_a(\vn)$ transforms like a link field \beq
U_a(\vn) \rightarrow G(\vn)U_a(\vn)G^\dagger(\vn + \hatbmu_a)\eeq
while the hermitian matrix $p_a(\vn)$ transforms like a scalar field \beq
p_a(\vn) \rightarrow G(\vn) p_a(\vn)G^\dagger(\vn).\eeq
 Thus in this language we can identify the $p_a(\vn)$ with the scalar field fluctuations
$B_a(\vn)$. The mass term then becomes
\beq
S_M = \mu^2 \sum_\vn \frac{1}{N^2} \Big[\frac{2}{\sqrt{N}} B^0_a(\vn)+ \sum_{A=1}^N (B^A_a(\vn))^2\Big]^2~.
\eeq

From this expression it is straightforward to see that the fluctuations of the scalar trace mode are governed by a quadratic potential while the traceless scalar field fluctuations feel only a quartic potential. Thus, if we keep $\mu\equiv\mu a$ fixed  as $a\to 0$ the trace
mode will acquire an infinite mass in the continuum limit and hence fluctuations
of the trace more around its vacuum expectation value will be completely
suppressed in that limit. In the same limit the presence of the quartic potential for the 
traceless Cartan generators is sufficient to regulate possible
infrared problems associated with the flat directions of the $SU(N)$ sector. 
Finally, once the continuum limit is attained,
we can restore supersymmetry by taking the final limit $\mu \to 0$.

Notice that the fact that this potential term selects out preferentially the trace mode of the scalars is trivially obvious if we adopt the exponential parametrization of the complexified gauge links
since in that case we can identify $I+p_a$ with $e^{iB_a}$.

\section{Simulation Results}
\label{sec:sim-results}

As noted previously, we have rescaled all lattice fields by powers of the lattice spacing to make them dimensionless. This leads to an overall dimensionless coupling parameter of the form $N/(2\lambda a^2)$, where $a = \beta/T$ is the lattice spacing, $\beta$ is the physical extent of the lattice in the Euclidean time direction and $T$ is the number of lattice sites in the time-direction. Thus, the lattice coupling is
\beq
\kappa = \frac{NL^2}{2\lambda\beta^2}~,
\eeq
for the symmetric two-dimensional lattice where the spatial length $L = T$\footnote{Notice that this coupling multiples {\it all} terms in the bosonic action including those associated with the scalar potential.}. Note that $\lambda\beta^2$ is the dimensionless physical `t Hooft coupling in units of the area. In our simulations\footnote{See \cite{Catterall:2011ce} for the details of the code we used to simulate these theories.}, the continuum limit can be approached by fixing $\lambda\beta^{2}$ and $N$ and increasing the number of lattice points $L\rightarrow\infty$. 
In practice we fix the value of $\beta = 1$ and vary $\lambda$.
We have taken three different values for this coupling  $\lambda= 0.5, 1.0, 2.0$ and lattice sizes ranging from $L = 2, \cdots, 16$.  Systems with $U(N)$ gauge groups with $N = 2, 3$ and $4$ have been examined.

The simulations are performed using anti-periodic (thermal) boundary conditions for the fermions\footnote{This forbids exact zero modes that are otherwise present in the fermionic sector.}. An RHMC algorithm was used for the simulations as described in \cite{Catterall:2011ce}. The use of a GPU accelerated solver \cite{Galvez:2011cd} allowed us to reach larger lattices than have thus far been studied.

\subsection{$\cQ = 4$ Supersymmetries}

\begin{figure}
\hspace{-1.2cm}
\includegraphics[scale=0.82]{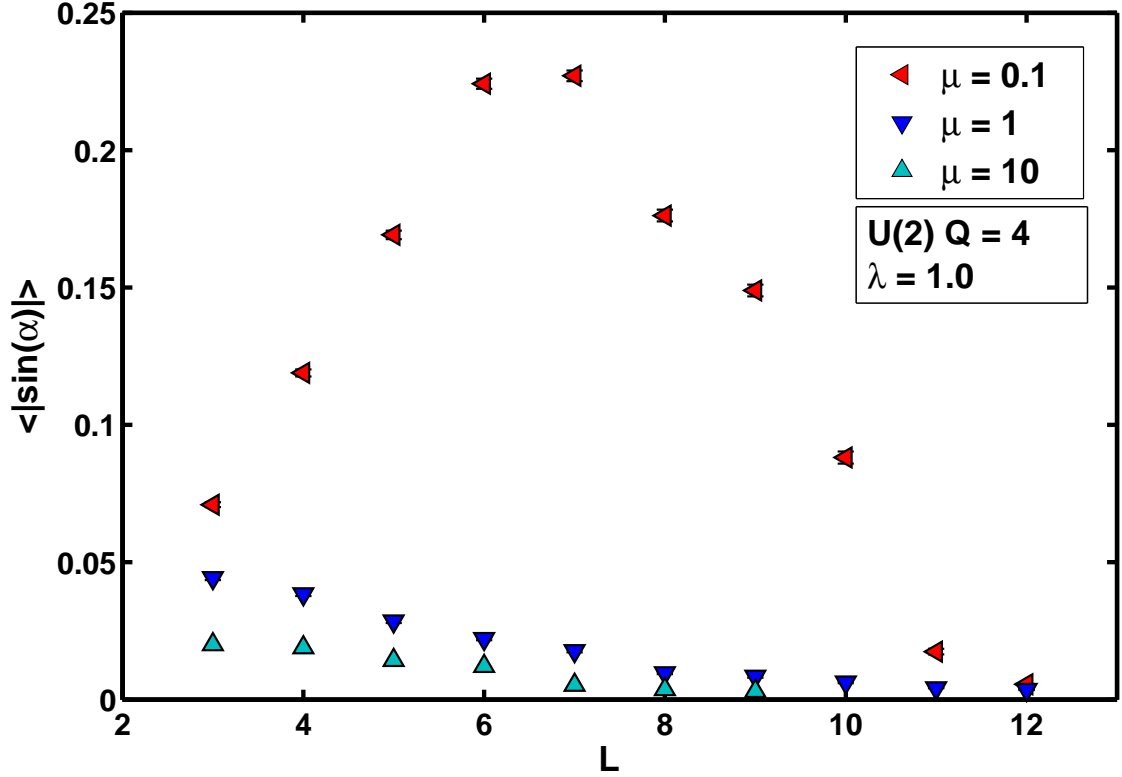}
\caption{$<|\sin{\alpha}|>$ for $\cQ=4$, U$(2)$ with $\mu=0.1, 1, 10$ }
\label{fig:U2_Q4_pfaffian}
\end{figure}

\begin{figure}
\hspace{-0.5cm}\includegraphics[scale=0.9]{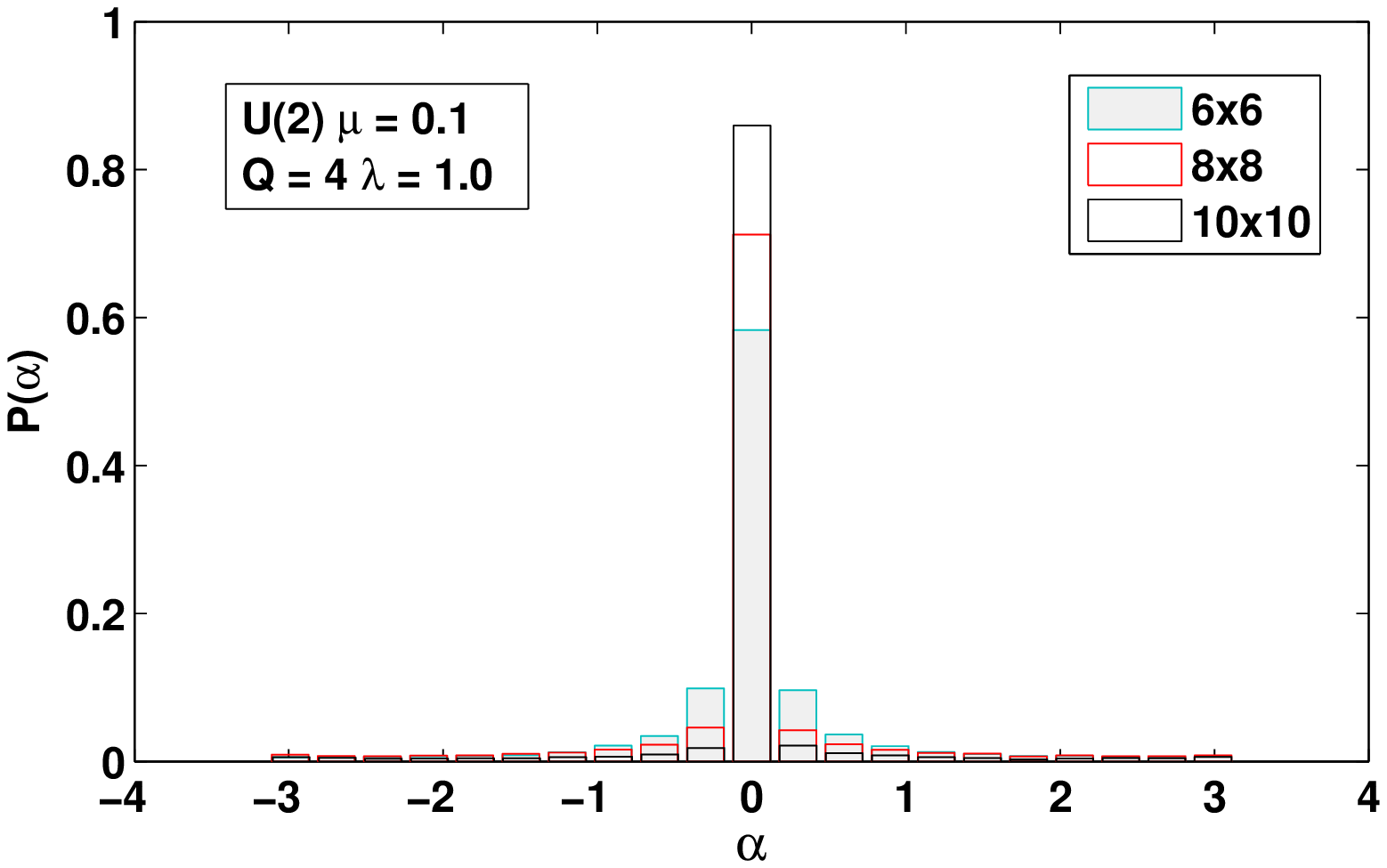}
\caption{Histogram for $\alpha$, with $\cQ=4$, U$(2)$, $\mu=0.1$ and volumes of 6x6, 8x8 and 10x10. \label{fig:Histogram_Q4}}
\end{figure}

In figure ~\ref{fig:U2_Q4_pfaffian}. we show results for the absolute value of the (sine of) the Pfaffian phase $|\sin{\alpha}|$ as a function of lattice size  $L = 1/a$ for the $\cQ = 4$ model with gauge group $U(2)$. The data corresponds to $\lambda=1$ but
similar results are obtained for $\lambda=0.5,2.0$ and larger numbers of
colors. Three values of $\mu$ are shown corresponding to $\mu=0.1$, $\mu=1.0$ and $\mu=10.0$.
While modest phase fluctuations are seen for small
lattices for the smallest value of $\mu$, 
we see that they disappear as the continuum limit is taken. As a practical
matter, these results make it clear that no re-weighting of observables is needed over much of the parameter space. This point is reinforced when we plot a histogram of the phase angle in figure \ref{fig:Histogram_Q4}. Clearly the angle fluctuations contract towards the origin as the
continuum limit is approached.

\begin{figure}
\hspace{-1.4cm}
\includegraphics[scale=0.82]{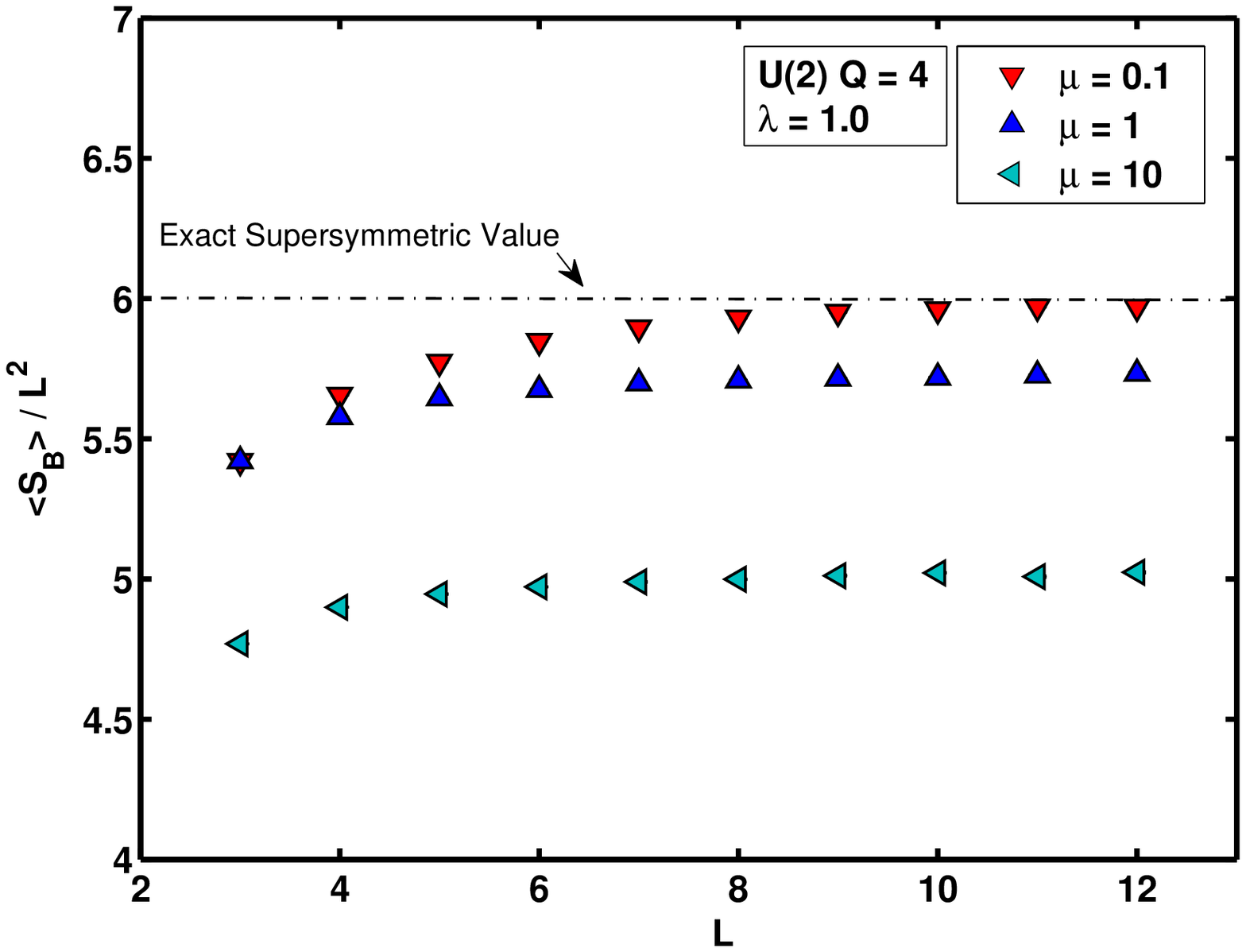}
\caption{$<\kappa S_B>$ for $\cQ=4$, U$(2)$ and $\mu=0.1, 1, 10$ }
\label{fig:Q4_ActionvsL}
\end{figure}

To check for the restoration of supersymmetry in the continuum limit and
as the scalar potential is sent to zero, we
show in  figure~\ref{fig:Q4_ActionvsL}.
a plot of the bosonic action density vs lattice size $L$. While the curves
plateau for large $L$ indicating a well defined continuum limit
it is clear that in general supersymmetry is broken there. Indeed, the exact value of
the bosonic action which is shown by the dotted line in the plot
can be computed using a simple $\cQ$ Ward identity and yields  \cite{Catterall:2008dv}
\beq
<\frac{1}{L^2}\kappa S_B>=\frac{3}{2}N_G\eeq
It should be clear from the plot that
the measured action indeed approaches this supersymmetric value if the subsequent limit $\mu\to 0$ is taken\footnote{Actually strictly we only expect this as $\beta\equiv \lambda\to\infty$ and thermal effects are suppressed. These appear to be already small for $\lambda=1$ in
this theory}. Thus the regulating procedure we have described does indeed provide a well defined
procedure for studying the supersymmetric lattice theory.

\begin{figure}
\hspace{-1.6cm}\includegraphics[scale=0.83]{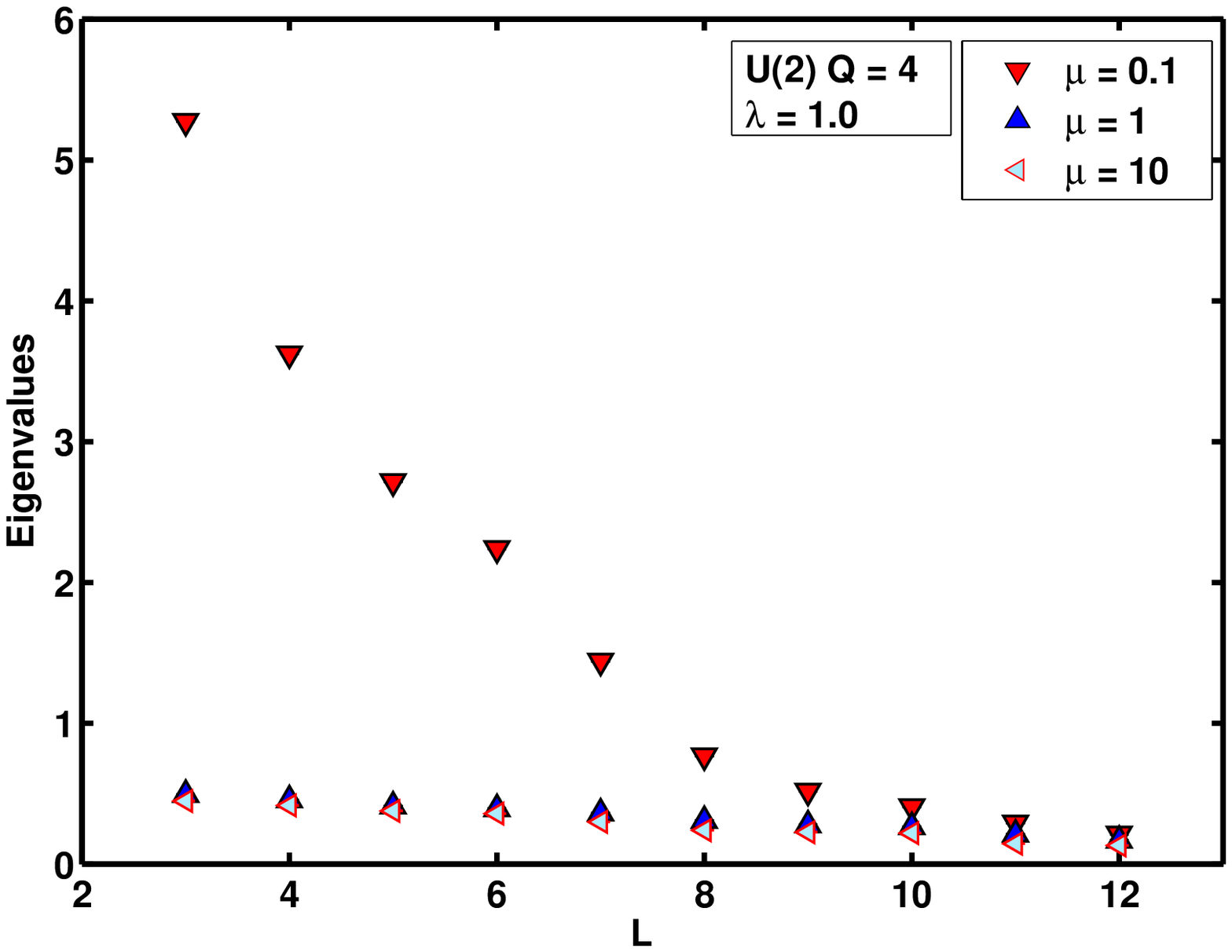}
\caption{Ensemble average for the eigenvalues of the  $(\cU_a^\dagger \cU_a-1)$ operator for $\cQ=4$, U$(2)$ and $\mu=0.1, 1, 10$ }
\label{fig:scalareigs}
\end{figure}

Finally, to reassure ourselves that $L\to\infty$ indeed corresponds to a continuum limit, figure \ref{fig:scalareigs}.
shows a plot of the expectation value of the maximal
eigenvalue of the operator $(\cU_a^\dagger \cU_a-1)$ averaged over the lattice
as a function of $L$ for $\lambda = 1$.
To leading order, this expression yields the largest scalar field eigenvalue in units of the lattice
spacing. Reassuringly we see that the eigenvalue indeed approaches zero as $L\to\infty$ corresponding
to a vanishing lattice spacing.

\subsection{$\cQ = 16$ Supersymmetries}

\begin{figure}
\hspace{-1.3cm}\includegraphics[scale=0.9]{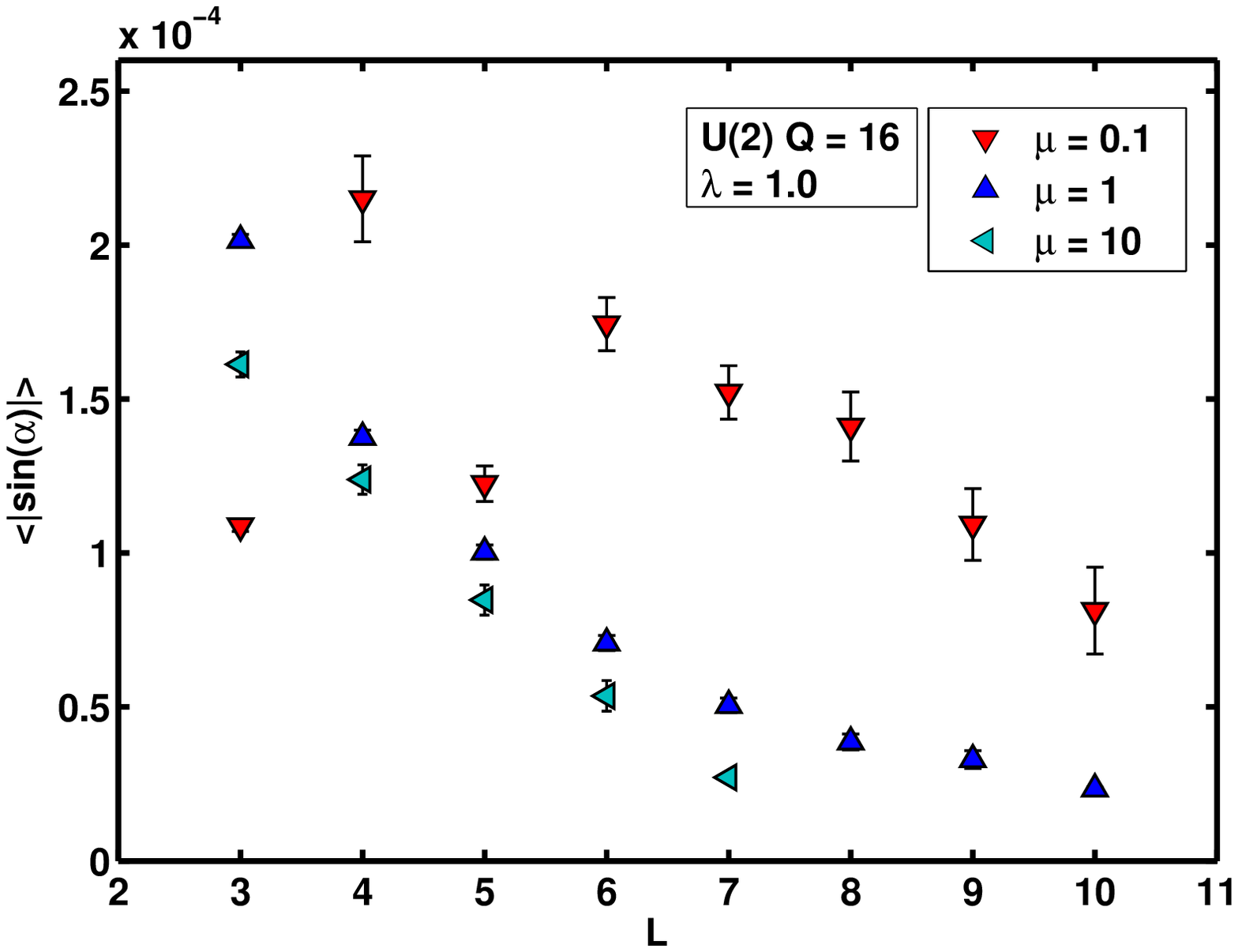}
\caption{$<|\sin{\alpha}|>$ for $\cQ=16$, U$(2)$ and $\mu=0.1, 1, 10$ \label{fig:Q16_pfaffian.eps}}
\end{figure}

\begin{figure}
\hspace{-0.35cm}\includegraphics[scale=0.9]{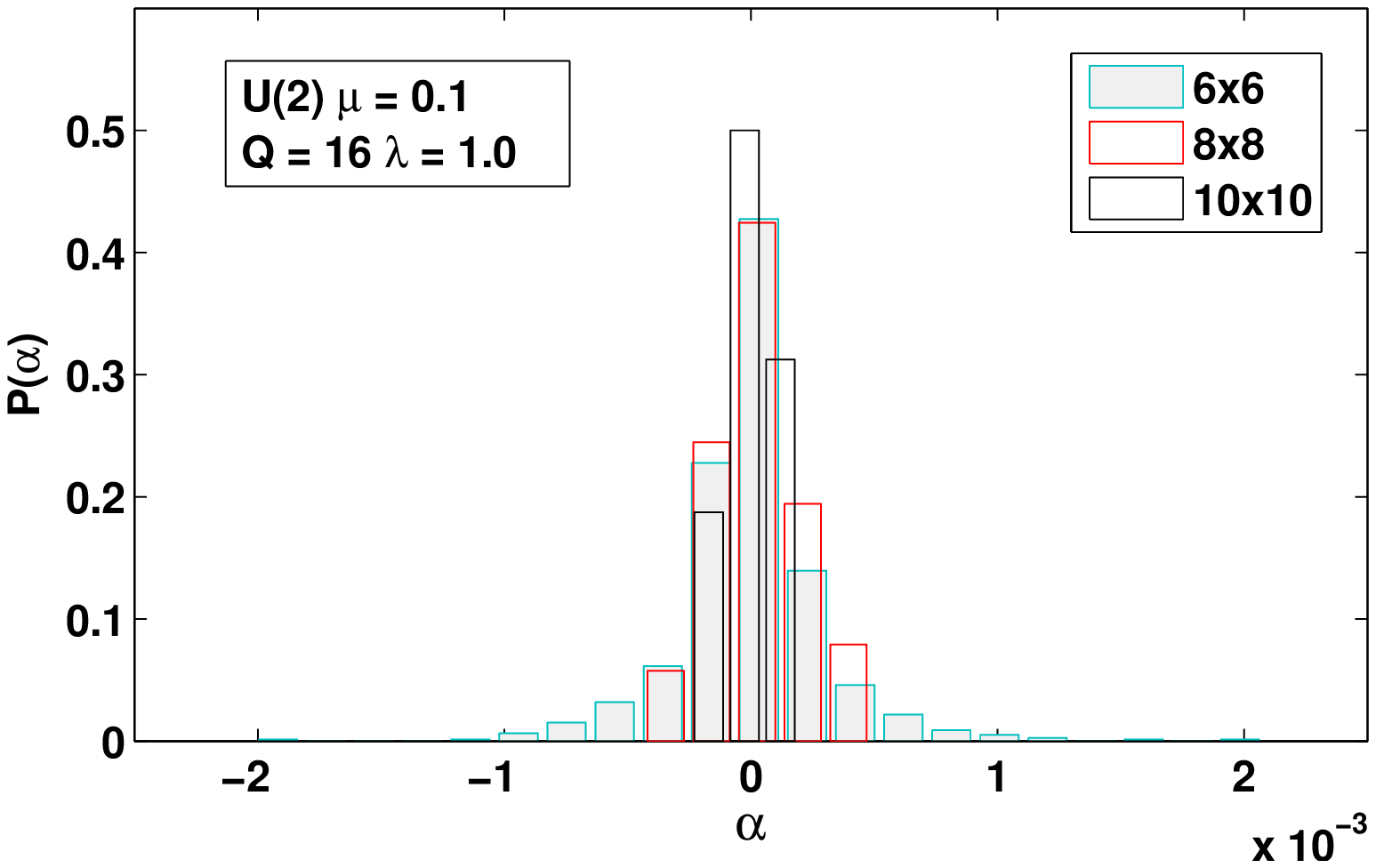}
\caption{Histogram for $\alpha$, with $\cQ=16$, U$(2)$, $\mu=0.1$ for volumes of 6x6, 8x8, and 10x10. \label{fig:Histogram_Q16}}
\end{figure}

The results for the absolute value of the (sine of) the Pfaffian phase for the $\cQ = 16$ supercharge model with U(2) gauge group
in two dimensions are shown in figure.~\ref{fig:Q16_pfaffian.eps}.
As for the $\cQ=4$ case, we see that the average Pfaffian phase is small and decreases with $L$. Indeed, the magnitude of
these angular fluctuations are $O(10^{-4})$ for all $L$ and $\mu$ - {\it much} smaller
than that observed for $\cQ=4$. Thus, even on the coarsest lattice
and smallest $\mu$, there is clearly no practical sign problem and certainly no sign problem in the continuum limit. Again, this picture is reinforced by looking at a histogram of the phase angle $\alpha$ as seen in figure \ref{fig:Histogram_Q16}. 

\begin{figure}
\hspace{-1.8cm}\includegraphics[scale=0.83]{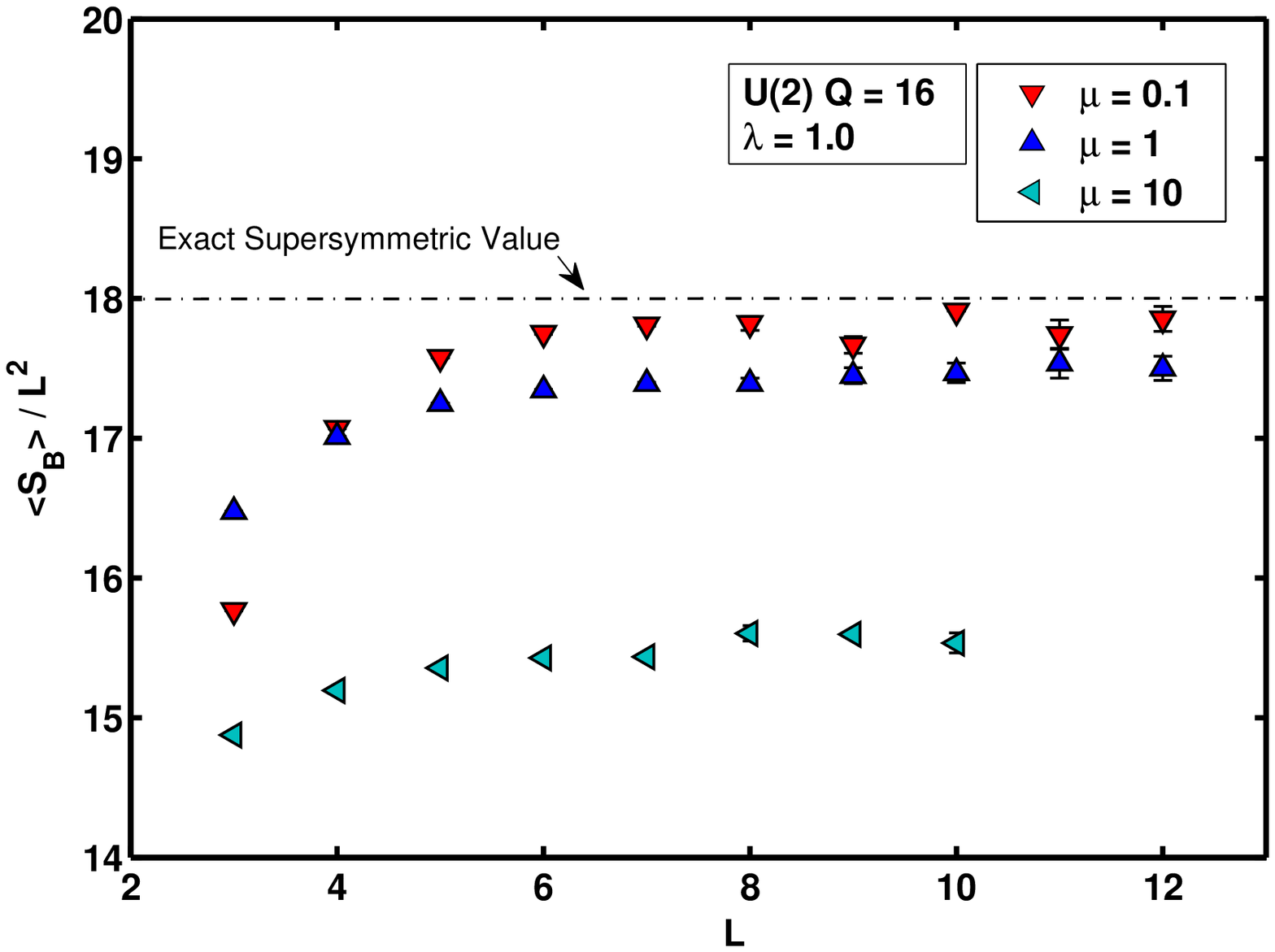}
\caption{$<S_B>$ for $\cQ=16$, U$(2)$ and $\mu=0.1, 1, 10$ \label{fig:Q16_ActionvsL}}
\end{figure}

\begin{figure}
\hspace{-1.6cm}\includegraphics[scale=0.83]{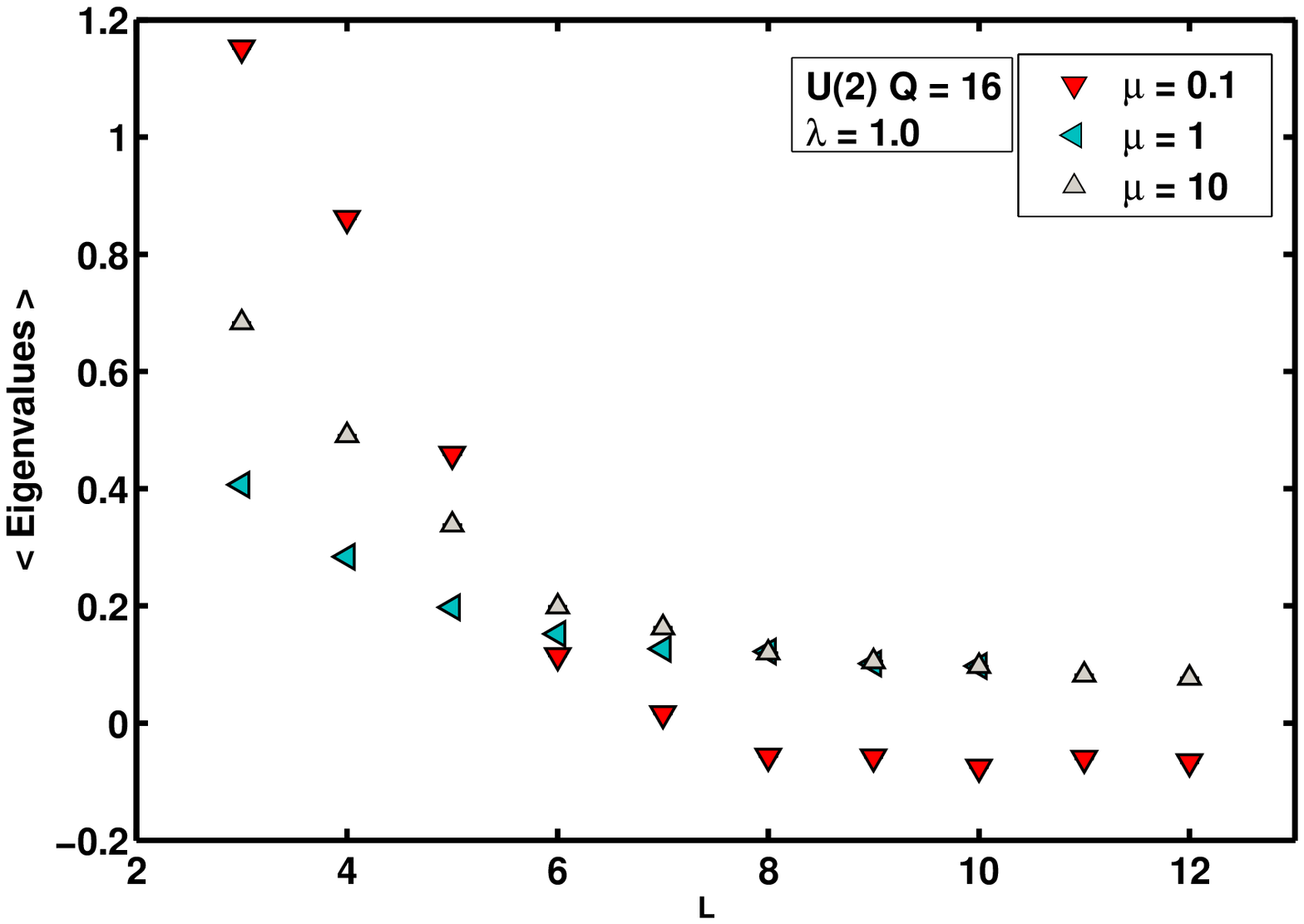}
\caption{Ensemble averaged eigenvalues for the  $(\cU_a^\dagger \cU_a-1)$ operator for $\cQ=16$ with U$(2)$ and $\mu=0.1, 1, 10$ \label{fig:Q16_EigenvsL}}
\end{figure}

The corresponding plot of the expectation value of the bosonic action vs lattice size $L$ is shown in figure~\ref{fig:Q16_ActionvsL}.
In the case of the $\cQ=16$ model the exact expression for the bosonic action is given by
\beq
<\frac{1}{L^2}\kappa S_B>=\frac{9}{2}N_G
\eeq
The data shown in this plot allow us to conclude that
 a well defined continuum limit exists for non-zero
$\mu$ and furthermore, $\cQ$-supersymmetry can
be restored
by subsequently sending the parameter $\mu\to 0$.
As a final cross check that the limit $L\to\infty$ indeed
corresponds to a true  continuum limit, we have again examined the the behavior of the 
maximal eigenvalue of $\cU^\dagger_a\cU-I$  as $L\to\infty$. 
The result is shown in figure~\ref{fig:Q16_EigenvsL}. and is consistent with a vanishing
lattice spacing in this limit.

\begin{figure}
\hspace{-1cm}\includegraphics[scale=0.93]{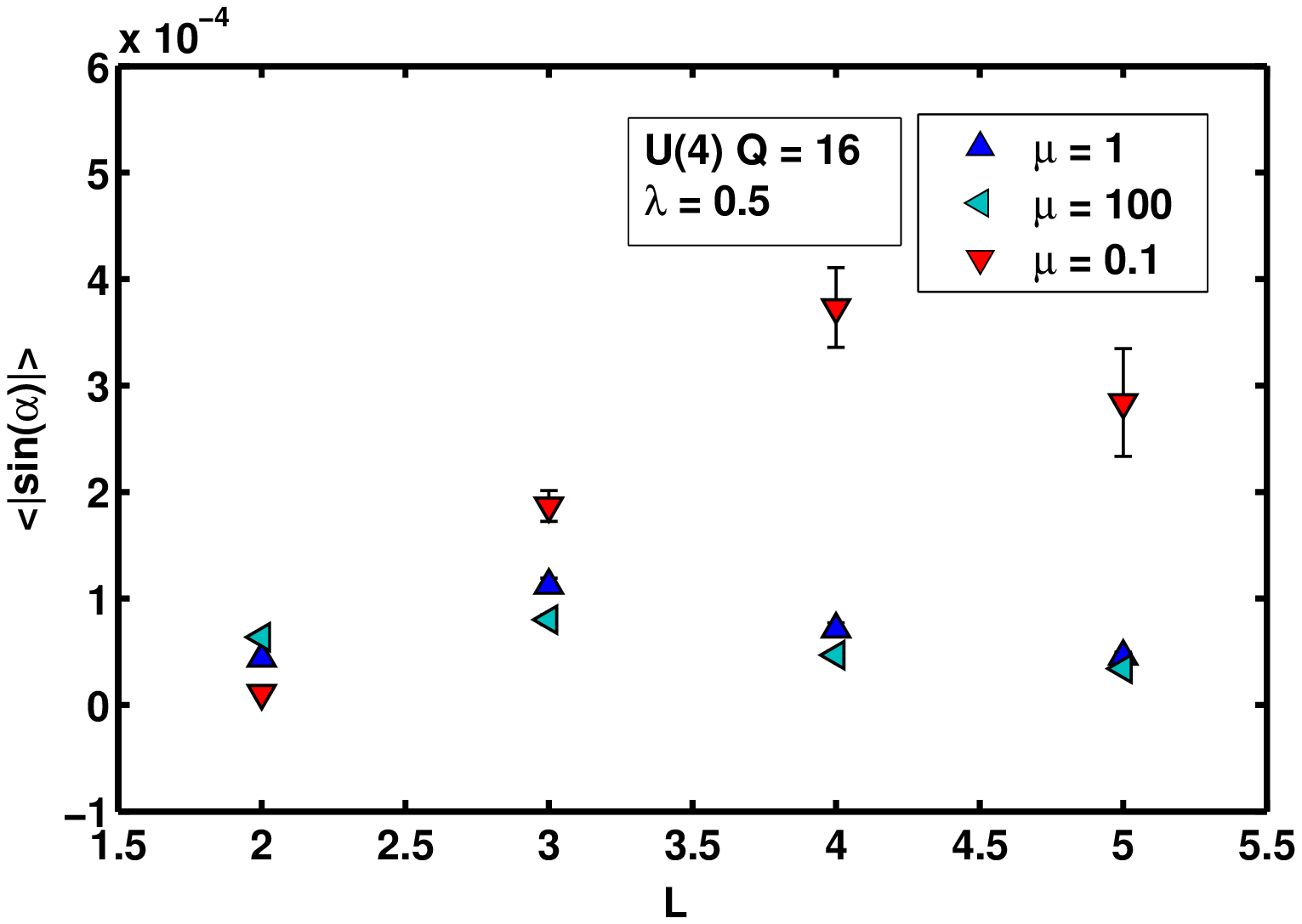}
\caption{$<|\sin{\alpha}|>$ for $\cQ=16$, U$(4)$ and $\mu=0.1, 1, 100$ \label{fig:U4_Q16_APBC_Pfaf}}
\end{figure}

These results generalize to large numbers of colors as can be seen in figure~\ref{fig:U4_Q16_APBC_Pfaf}. where we plot the expectation value of
the absolute value of the sine of the Pfaffian phase for the case
of the U(4) group. Notice that the Pfaffian can be proven real in the limit that the $\cQ=16$ theory
is reduced to zero dimensions
for two and three colors
so that  it is necessary to examine the $U(4)$ case to be sure of seeing truly
generic behavior.

Nevertheless we see that U(4) looks qualitatively the same  as  for U(2). In fact  the fluctuations in the phase angle that we observe are even
{\it smaller} than those seen for the U(2) theory. This again indicates that this theory exhibits no sign problem even on small lattices and certainly in the continuum limit.

\begin{figure}
\hspace{-1.6cm}\includegraphics[scale=0.83]{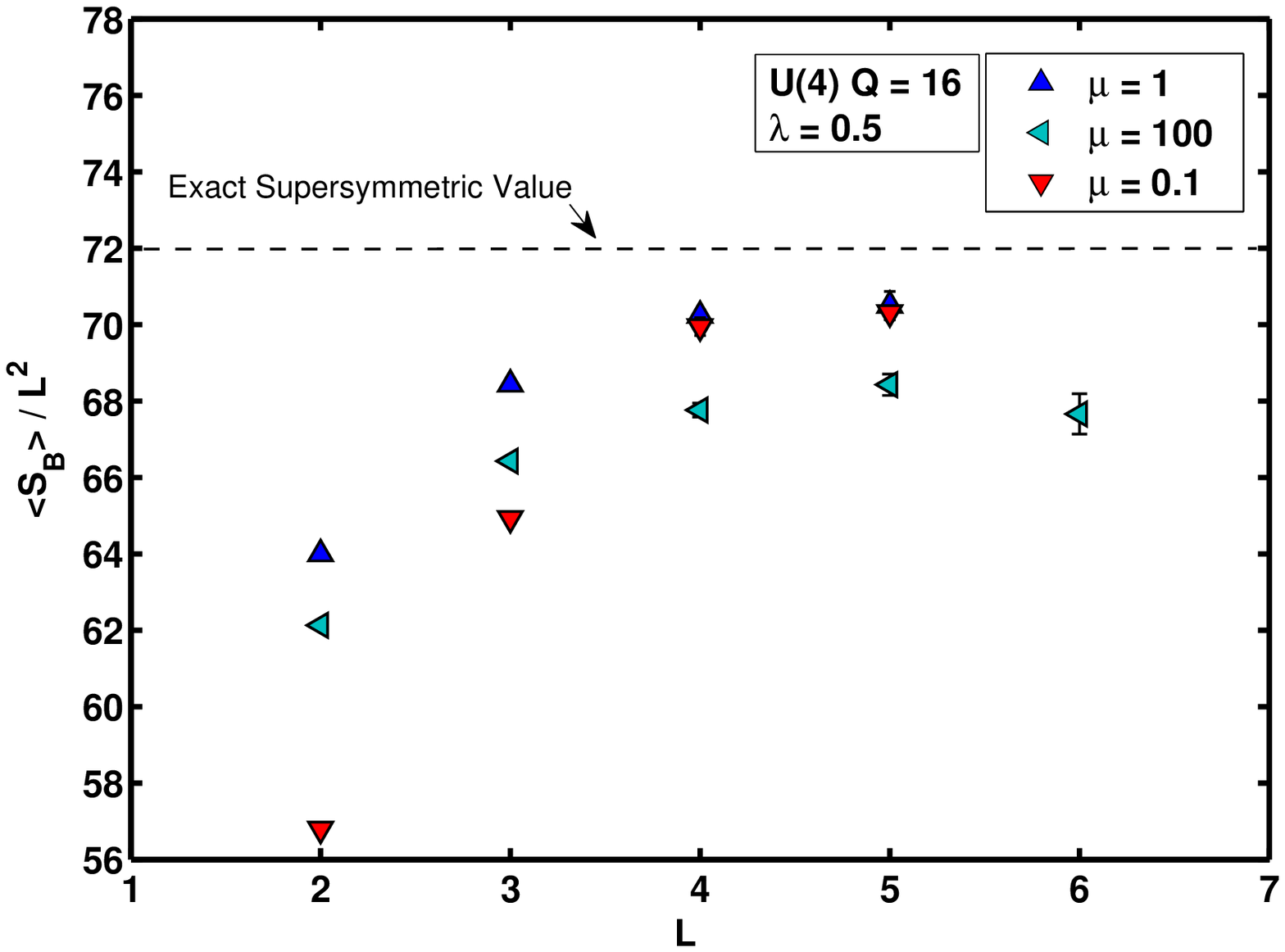}
\caption{$<S_B>/L^2$ for $\cQ=16$, U$(4)$ and $\mu=0.1, 1, 100$ \label{fig:U4_Q16_APBC_Action}}
\end{figure}

\begin{figure}
\hspace{-1.6cm}\includegraphics[scale=0.83]{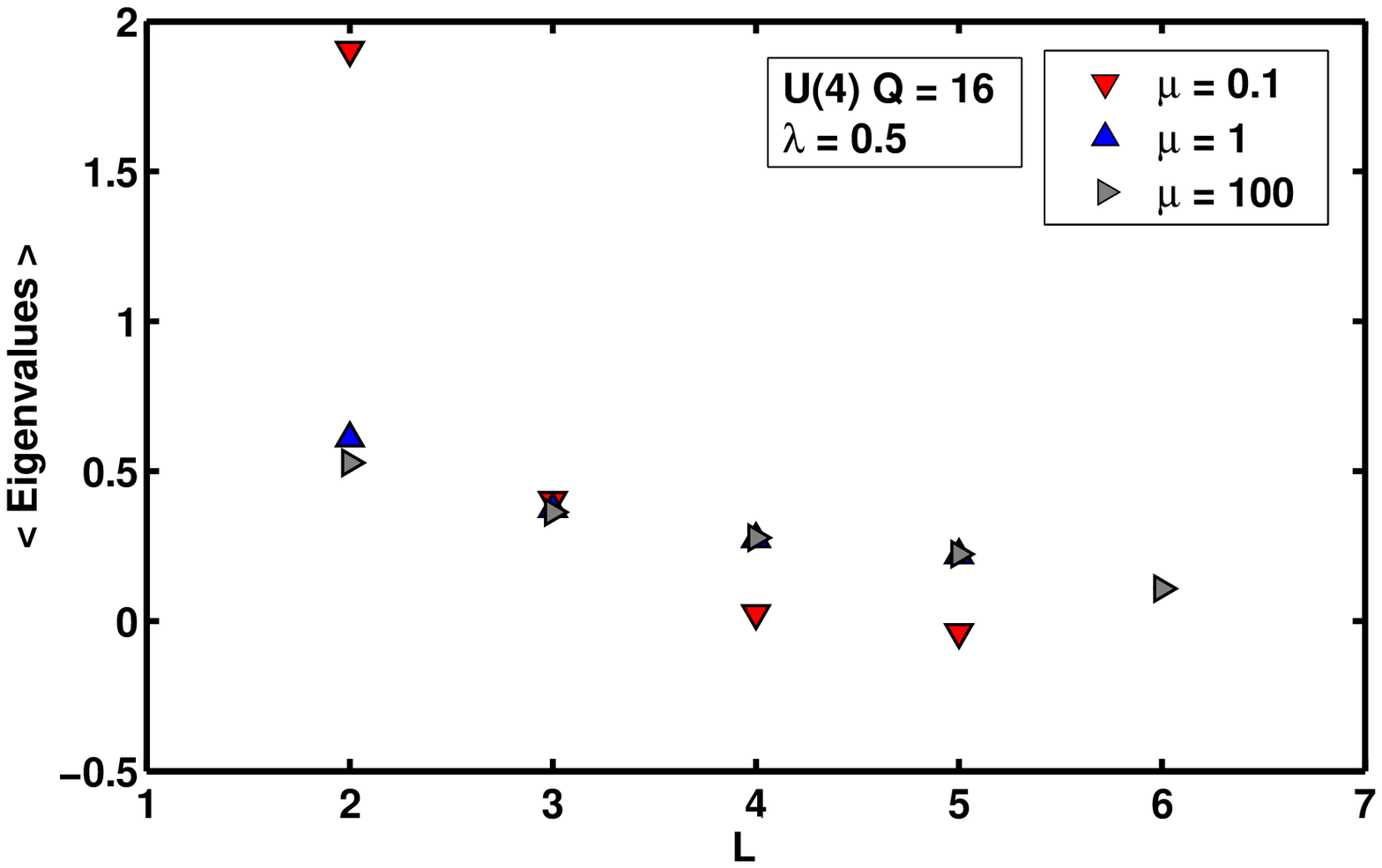}
\caption{Ensemble averaged eigenvalues for the $(\cU_a^\dagger \cU_a-1)$ operator for $\cQ=16$, U$(4)$ and $\mu=0.1, 1, 100$ \label{fig:U4_Q16_APBC_Eigen}}
\end{figure}

The plot of the bosonic action for U(4) is shown in figure~\ref{fig:U4_Q16_APBC_Action}. While the largest lattice we have been able to
simulate thus far is rather too small to get a good continuum limit the measured bosonic action
is nevertheless within a percent or so of the exact value expected on the basis of
$\cQ$-supersymmetry. The scalar field fluctuations also decrease toward zero as the
number of lattice points increase as shown in  figure \ref{fig:U4_Q16_APBC_Eigen}. 

It is at first sight rather remarkable that the observed Pfaffian phase fluctuations
are small in the $\cQ=16$ theory given that the Pfaffian is certainly complex when
evaluated  on a generic
set of background scalar and gauge fields. It appears to be a consequence of very specific
dynamics in the  theory which ensure that only certain special regions of
field space are important in the path integral. Of course the continuum theory does possess very special dynamics; 
for example the twisted supersymmetry ensures that
the torus partition function $Z$ is a topological invariant. 
One immediate consequence of this is that
$Z$ may be computed exactly at one loop where Marcus has argued that
it simply reduces to an unsigned sum over isolated points in the moduli space of flat
complexified connections up to complex gauge transformations  \cite{Marcus:1995mq}.
Furthermore, much of this structure survives in the {\it lattice} theory; the full
partition function {\it including} any Pfaffian phase 
may be calculated exactly at one loop. As in the continuum theory there is
a perfect cancellation of contributions from  fermons and
bosons and the final result is real \cite{Catterall:2011pd}. Of course this does not mean that simulations
at finite gauge coupling should not suffer from 
sign problems but certainly makes it less likely. More prosaically, it is  easy to see
that the Pfaffian is real positive if the lattice scalar fields are set to zero - and this is what effectively
happens in the continuum limit as a result of the scalar potential that we use to control the
vacuum expectation value and fluctuations of the trace mode.
\section{Conclusions}
 
We have performed numerical simulations of the four and sixteen supercharge lattice SYM theories in two dimensions to investigate the occurrence of a sign problem in these theories. In contrast to the usual situation in lattice gauge theory, we utilize a non compact parameterization of the gauge fields in which the lattice fields are expanded on the algebra of the group. While such a scheme would ordinarily break lattice gauge invariance we show that in the case of these twisted supersymmetric models this preserves gauge symmetry since the models in question are formulated in terms of a complexified gauge field valued in $U(N)$. The correct continuum limit is then ensured by adding an appropriate gauge invariant potential term which picks out a non-zero vacuum expectation value for the trace mode of the scalar fields in the continuum limit. We argue that the effects of this potential on the remaining traceless modes can be subsequently removed by sending the potential to zero {\it after} the continuum limit is taken.

We have examined both supersymmetric  theories for several values of  the dimensionless 't Hooft coupling $\lambda \beta^2$ and for gauge groups $U(2)$, $U(3)$ and $U(4)$. We take a careful continuum limit by simulating the theories over a range of lattice size $L = 2-14$. In both cases we see that the average Pfaffian phase goes to zero for a fixed gauge invariant potential as the continuum limit is taken. We also examine the subsequent limit in which the potential is removed and show evidence that supersymmetry is restored. While the absence of a sign problem is not surprising in the $\cQ=4$ case (where one can prove the Pfaffian reduces to a real positive definite determinant in the continuum limit) it is much more non trivial matter in the $\cQ=16$ supercharge case. In that case the Pfaffian evaluated on a generic background is complex even in the continuum limit. Nevertheless, we observe that the Pfaffian phase is
small and decreases to zero as the continuum limit is taken. Indeed, in practice it is sufficiently small even on coarse lattices that there is no need to use a reweighting procedure to compute expectation values of observables. The analysis of the $\cQ=16$ model is complicated by the fact that the $U(2)$ and $U(3)$ theories exhibit some special properties since in the matrix model limit they are real positive definite and real respectively. Nevertheless, the pattern we observe for the $U(4)$ group is similar to that seen for the smaller groups and the trend supports the conjecture that the sign problem is absent in the continuum limit.

These results thus help to strengthen the case that there may be no sign problem for the $\cQ = 16$ theory in four dimensions and hence no a priori barrier to numerical studies of this
theory.

\begin{acknowledgments}
This work was supported by the U.S. Department of Energy grant under contract no. DE-FG02-85ER40237 and Science Foundation Ireland grant 08/RFP/PHY1462. Simulations were performed using USQCD resources at Fermilab. The authors would like to acknowledge valuable conversations with Joaquin Drut and Robert Wells. AJ's work is also supported in part by the LDRD program at the Los Alamos National Laboratory.
\end{acknowledgments}

\bibliographystyle{JHEP}

\end{document}